\documentclass[paper=a4,fontsize=11pt,numbers=noendperiod]{scrartcl}
\usepackage[T1]{fontenc}
\usepackage[utf8]{inputenc}
\usepackage[english]{babel}
\usepackage[protrusion=true,expansion=true]{microtype}
\usepackage{amsmath,amsfonts,amsthm,amssymb}
\usepackage[pdftex]{graphicx}
\usepackage[dvipsnames]{xcolor}
\usepackage[pdftex,colorlinks]{hyperref}
\usepackage{cleveref}
\usepackage[autostyle]{csquotes}
\usepackage[bf]{caption}
\usepackage{subcaption}
\usepackage{tikz}
\usepackage{xifthen}
\usepackage{pgfplots}
\usepackage{floatrow}
\usepackage{float}
\usepackage[section,notindex,nottoc]{tocbibind}
\usepackage[backend=bibtex,style=numeric,sorting=ynt,citestyle=numeric,maxbibnames=100,bibencoding=utf8]{biblatex}
\usepackage{etoolbox}
\usepackage[tocgraduated]{tocstyle}
\usepackage{blindtext}
\usepackage{setspace}
\usepackage{pbox}
\usepackage{titlesec}
\usepackage{appendix}
\usepackage{pgf}
\usepackage{mathtools}
\usepackage{xifthen}
\usepackage{pgfplots}
\usepackage{rotating}
\usepackage{breakurl}
\usepackage{floatrow}
\usepackage{authblk}

\usetikzlibrary{calc}
\usetikzlibrary{calc,positioning,shadows,arrows}
\tikzstyle{process} = [rectangle, minimum width=3cm, minimum height=1cm, text centered, draw=black, fill=blue!10]
\tikzstyle{arrow} = [thick,->,>=stealth]
\usetikzlibrary{%
  arrows,%
  calc,%
  fit,%
  patterns,%
  plotmarks,%
  shapes.geometric,%
  shapes.misc,%
  shapes.symbols,%
  shapes.arrows,%
  shapes.callouts,%
  backgrounds,%
  chains,%
  topaths,%
  trees,%
  petri,%
  matrix,%
  folding,%
  fadings,%
  through,%
  positioning,%
  scopes,%
  decorations.fractals,%
  decorations.shapes,%
  decorations.text,%
  decorations.pathmorphing,%
  decorations.pathreplacing,%
  decorations.footprints,%
  decorations.markings,%
  shadows}

\DeclareMathOperator*{\vectorize}{vec}

\newcommand{\bfcaption}[2]{\caption[#1]{\textbf{#1:} #2}}

\newcommand{\lnorm}[1]{\( \ell_{#1} \)}

\KOMAoption{listof}{toc,leveldown}

\title{Sparse convolutional coding for neuronal ensemble identification}

\usetocstyle{allwithdot}

\AtEveryCitekey{\clearfield{url}}

\newcommand\myshade{85}
\colorlet{mylinkcolor}{violet}
\colorlet{mycitecolor}{YellowOrange}
\colorlet{myurlcolor}{Aquamarine}
\definecolor{mytitle}{HTML}{ad3136}

\hypersetup{
  linkcolor  = mylinkcolor!\myshade!black,
  citecolor  = mycitecolor!\myshade!black,
  urlcolor   = myurlcolor!\myshade!black,
  colorlinks = true,
}

\numberwithin{equation}{section}
\numberwithin{figure}{section}
\numberwithin{table}{section}

\author[1]{\href{mailto:sven.peter@iwr.uni-heidelberg.de}{Sven Peter}}
\author[2]{\href{mailto:daniel.durstewitz@zi-mannheim.de}{Daniel Durstewitz}}
\author[1]{\href{mailto:ferran.diego@iwr.uni-heidelberg.de}{Ferran Diego}}
\author[1]{\href{mailto:fred.hamprecht@iwr.uni-heidelberg.de}{Fred A. Hamprecht}}
\affil[1]{Heidelberg Collaboratory for Image Processing, Heidelberg University}
\affil[2]{Department of Theoretical Neuroscience, Bernstein Center for Computational Neuroscience, Central Institute of Mental Health, Medical Faculty Mannheim of Heidelberg University}
\begin{document}

\maketitle

\begin{abstract}
\textit{Cell ensembles}, originally proposed by Donald Hebb in 1949, are subsets of synchronously firing neurons and proposed to explain basic firing behavior in the brain. Despite having been studied for many years no conclusive evidence has been presented yet for their existence and involvement in information processing such that their identification is still a topic of modern research, especially since simultaneous recordings of large neuronal population have become possible in the past three decades. These large recordings pose a challenge for methods allowing to identify individual neurons forming cell ensembles and their time course of activity inside the vast amounts of spikes recorded. Related work so far focused on the identification of purely simultaneously firing neurons using techniques such as Principal Component Analysis. In this paper we propose a new algorithm based on sparse convolution coding which is also able to find ensembles with temporal structure. Application of our algorithm to synthetically generated datasets shows that it outperforms previous work and is able to accurately identify temporal cell ensembles even when those contain overlapping neurons or when strong background noise is present.

\end{abstract}

\section{Introduction}
\textit{Cell ensembles} (or synonymously \textit{cell assemblies} or \textit{cortical motifs}) were originally proposed by Hebb \cite{hebb1949organization} as subsets of synchronously firing neurons to explain brain activity underlying complex behaviors. 
Multiple studies have been done to find evidence for or against the neuronal ensemble hypothesis on datasets recorded from different areas of the brain inside different animals but no clear answer has been found yet \cite{singer1993synchronization,marr1991simple,pastalkova2008internally,nicolelis1997hebb,mokeichev2007stochastic}. 

\begin{figure}[h]
    \centering
    \subcaptionbox{Synchronously firing neurons\label{fig:ensemble-spatial}}{
    \begin{tikzpicture}[scale=1.6]
      \foreach \t in {0, 0.05, 0.1, ..., 2.5} {

        \pgfmathsetmacro{\rnd}{random(8)}

        \ifthenelse{\NOT 1 = \rnd}{
          \pgfmathsetmacro{\ctr}{0}
          \pgfplotsforeachungrouped \neuron in {0, 0.20, 0.40, 0.60} {
            \pgfmathsetmacro{\rnd}{random(6)}
            \ifthenelse{\NOT 1 = \rnd}{}{
              \pgfmathsetmacro{\ctr}{\ctr + 1};
              \ifthenelse{\NOT 4 = \ctr}{
                \fill [black] (\t + 0.0,\neuron + 0.02) rectangle (\t + 0.02,\neuron+0.2);
              }{};
            }
          }
        }{
          \foreach \neuron in {0, 0.20, 0.40, 0.60} {
            \fill [red!50] (\t + 0.0,\neuron + 0.02) rectangle (\t + 0.02,\neuron+0.2);
          }   
        }
      }
    \end{tikzpicture}}\hfill%
    \subcaptionbox{Synfire chain\label{fig:ensemble-synfire}}{
    \begin{tikzpicture}[scale=1.6]
      \foreach \majort in {0, 0.25, 0.5, ..., 2.5} {
        \pgfmathsetmacro{\rnd}{mod(\majort*4,3)}

        \ifthenelse{\NOT 0 = \rnd}{
          \foreach \minort in {0, 0.05, 0.1, 0.15, 0.2} {
            \pgfmathsetmacro{\t}{\majort + \minort}
            \foreach \neuron in {0, 0.20, 0.40, 0.60} {
              \pgfmathsetmacro{\rnd}{random(6)}
              \ifthenelse{\NOT 1 = \rnd}{}{
              \fill [black] (\t + 0.0,\neuron + 0.02) rectangle (\t + 0.02,\neuron+0.2);
              }
            }
          }
        }{
          \fill [red!50] (\majort,0.02) rectangle (\majort + 0.02, 0.2);

          \fill [red!50] (\majort+0.05,0.20+0.02) rectangle (\majort + 0.05 + 0.02, 0.2+0.2);

          \fill [red!50] (\majort+0.1,0.40+0.02) rectangle (\majort + 0.1 + 0.02, 0.2+0.4);

          \fill [red!50] (\majort+0.15,0.60+0.02) rectangle (\majort + 0.15 + 0.02, 0.2+0.6);
        }
      }
    \end{tikzpicture}}\hfill%
    \subcaptionbox{Temporal motif\label{fig:ensemble-lag}}{
    \begin{tikzpicture}[scale=1.6]
      \foreach \majort in {0, 0.25, 0.5, ..., 2.5} {
        \pgfmathsetmacro{\rnd}{mod(\majort*4,3)}

        \ifthenelse{\NOT 0 = \rnd}{
          \foreach \minort in {0, 0.05, 0.1, 0.15, 0.2} {
            \pgfmathsetmacro{\t}{\majort + \minort}
            \foreach \neuron in {0, 0.20, 0.40, 0.60} {
              \pgfmathsetmacro{\rnd}{random(6)}
              \ifthenelse{\NOT 1 = \rnd}{}{
              \fill [black] (\t + 0.0,\neuron + 0.02) rectangle (\t + 0.02,\neuron+0.2);
              }
            }
          }
        }{
          \fill [red!50] (\majort,0.02) rectangle (\majort + 0.02, 0.2);
          \fill [red!50] (\majort,0.20+0.02) rectangle (\majort + 0.02, 0.20 + 0.2);  
          \fill [red!50] (\majort,0.40+0.02) rectangle (\majort + 0.02, 0.40 + 0.2);  

          \fill [red!50] (\majort+0.05,0.02) rectangle (\majort + 0.05 + 0.02, 0.2);

          \fill [red!50] (\majort+0.1,0.20+0.02) rectangle (\majort + 0.1 + 0.02, 0.2+0.2);

          \fill [red!50] (\majort+0.15,0.00+0.02) rectangle (\majort + 0.15 + 0.02, 0.0+0.2);
          \fill [red!50] (\majort+0.15,0.60+0.02) rectangle (\majort + 0.15 + 0.02, 0.6+0.2);

          \fill [red!50] (\majort+0.2,0.00+0.02) rectangle (\majort + 0.2 + 0.02, 0.0+0.2);
          \fill [red!50] (\majort+0.2,0.40+0.02) rectangle (\majort + 0.2 + 0.02, 0.4+0.2);
          \fill [red!50] (\majort+0.2,0.60+0.02) rectangle (\majort + 0.2 + 0.02, 0.6+0.2);
        }
      }
    \end{tikzpicture}}

    \caption{\textbf{Synchronous and temporal motifs}: All three illustrations show four neurons spiking at various times. The spikes highlighted in red are part of a repeating motif. In (a) the pattern is just the synchronous activity of all neurons while the synfire chain in (b) and the motif in (c) also contain temporal structure, respectively}
\end{figure}
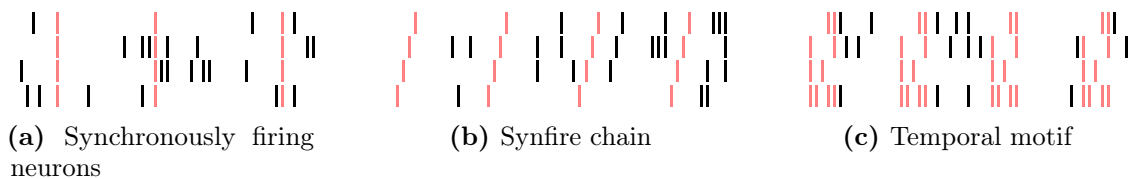

Especially in past two decades where it became possible to record large neuronal populations concurrently \cite{buzsaki2004large,stevenson2011advances,ahrens2013whole} methods such as Principal Component Analysis (PCA)  \cite{chapin1999principal} / Singular Value Decomposition (SVD) \cite{carrillo2015endogenous}, Independent Component Analysis (ICA) \cite{lopes2013detecting} or Non Negative Matrix Factorization (NMF) \cite{diego_13_learning} have been applied to identify neurons repeatedly firing at the same time (see figure \ref{fig:ensemble-spatial}) to find statistically significant ensembles and answer the question about their existence. While there are also more complex methods dealing with jitter in individual spike times \cite{Billeh201492} or the recurrence of motifs involving the ensembles themselves \cite{diego_13_learning,carrillo2015endogenous} more complex motifs such as synfire chains \cite{ikegaya2004synfire} (see figure \ref{fig:ensemble-synfire}) or motifs where only a single neuron is active at a certain time (see figure \ref{fig:ensemble-lag}) are still missed \cite{lopes2013detecting}.

With the increasing size of available datasets more powerful methods are required to not only handle low signal-to-noise levels but to also understand the formation and existence of ensembles with more complex temporal structures. In this paper we leverage sparsity constraints on neuronal activity to allow a simple and elegant mathematical formulation to identify ensembles completely unsupervised. We show that our proposed algorithm outperforms state-of-the-art methods on synthetic data.

\section{Related work}
\textcite{lopes2013detecting} provide an overview of using principal and independent component analysis (PCA and ICA) to identify ensembles. Both require an estimation of the number of ensembles $N_e$ at first which is based on the number of significant eigenvalues of the correlation matrix. Significance is either established using random matrix eigenvalue distribution theory \cite{marchenko1967distribution} or by shuffling the spike matrix to remove temporal correlation while preserving spike count distribution.

PCA, which has been used for a long time to track cell ensembles \cite{nicolelis1995sensorimotor}, computes the first $N_e$ principal components of the spike matrix and considers those to be the ensembles. Its biggest limitations are that two different ensemble patterns can be merged into a single component and that neurons shared between two ensembles are assigned lower weights than expected. Additionally negative values with no physical meaning are possible in the components. Recovering individual neurons which belong to a single ensemble is not reliably possible \cite{lopes2011neuronal,lopes2013detecting}.

ICA decomposes a multivariate signal into additive subcomponents with the assumption that these are non-Gaussian and statistically independent from each other \cite{comon1994independent}. When used to learn ensembles it overcomes some of the problems of PCA-based methods: Individual neuron-ensemble membership can be recovered easily and neurons belonging to multiple ensembles are also correctly identified \cite{lopes2013detecting}. Again negative values are possible in the identified patterns leading to interpretation problems. For synchronous patterns \textcite{lopes2013detecting} recommend to use this method since it provides the best estimate. Temporal structure is however not part of this model and patterns such as synfire chains cannot be identified. 

\textcite{carrillo2015endogenous} use singular value decomposition (SVD) to identify synchronously firing neurons. They identify directionally sensitive ensembles in the visual cortex of mice and are able to correlate their activity to the time during which external stimuli were presented. They are also able to identify ensembles sensitive to natural scenes and show repeating activations of those by applying graph theoretical methods to the already identified ensembles. Unlike our method however the number of ensembles has to be estimated at first and only synchronously spiking neurons are considered in the first step. While they can find activity spread over time these have to have at least two neurons firing synchronously in every bin, such that synfire chains would not be identified again.

\textcite{Billeh201492} tried to identify \textit{almost} synchronously firing neurons: Instead of relying on almost perfect synchronicity or on appropriate binning to ensure this condition they take into account that individual neurons belonging to a single ensemble will show a small jitter in their spiking time and used a dynamics-driven methodology based on the Markov Stability framework \cite{Delvenne20072010} to identify ensembles at multiple levels of granularity. While they are able to identify connections between neurons even if the activity is shifted in time they do not identify the exact motifs: Their result is the connectivity between neurons while we identify the connections and their temporal relation simultaneously instead.

\textcite{diego_13_learning} propose to use non-negative matrix factorization techniques to decompose binned spike matrix into multiple levels of synchronous patterns to identify a hierarchical structure of motifs. Again no temporal structure is taken into account and only neurons firing at the same time are considered.

Approaches based on one dimensional convolutional coding \cite{Szlam_10_convolutional} have been used to recover spike trains of individual neurons from recorded Calcium fluorescence sequences \cite{andilla2014sparse,pnevmatikakis_13_rank,pnevmatikakis_13_sparse,Vogelstein2009b}. In these models each neuron is however treated independently and they have not been extended to model relationships between neurons. In order to extract motifs a novel optimization approach is required to replace the one dimensional with two dimensional filters.

We propose to adapt the idea behind non negative convolutional matrix factorization \cite{o2006convolutive}, which has originally been developed to allow the extraction of motifs in audio processing, for the learning of motifs. This method only regularizes the activity of the learned motifs with a \lnorm{1} prior which is too weak to recover motifs in neuronal spike trains. Instead we propose a different optimization technique that allows to regularize the motifs with a \lnorm{1} and their activities with a much stronger \lnorm{0} prior. This allows for a simple and elegant formulation to learn complex motifs from recorded spike trains.

\section{Our method}
\begin{figure}[h]
  \begin{tikzpicture}
  \node[inner sep=0pt,label=below:{$\mathbf{Y}$}] (russell) at (0,4) {\includegraphics[height=40px]{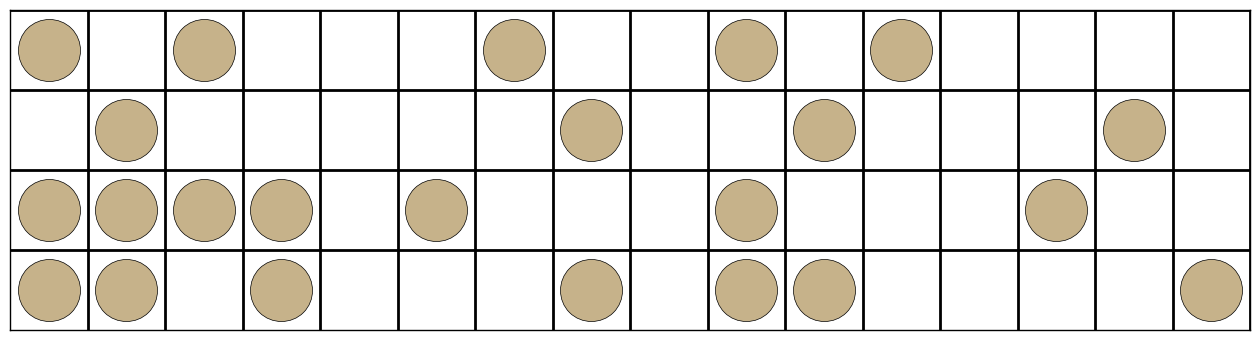}};
  \node[inner sep=0pt] (russell) at (3,4) {$=$};

  \node[inner sep=0pt,label=below:{$\mathbf{X}$}] (russell) at (6,4) {\includegraphics[height=40px]{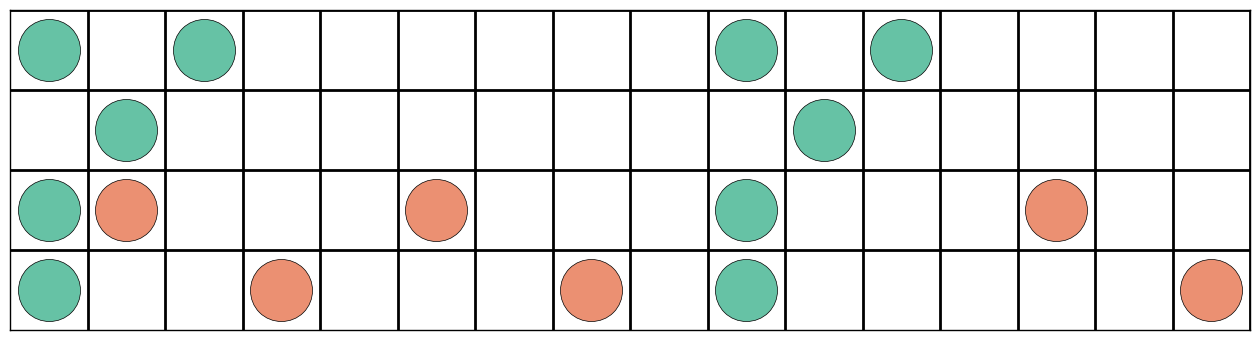}};
  \node[inner sep=0pt] (russell) at (6,2.5) {$+$};
  \node[inner sep=0pt,label=below:{$\mathbf{N}$}] (russell) at (6,1.5) {\includegraphics[height=40px]{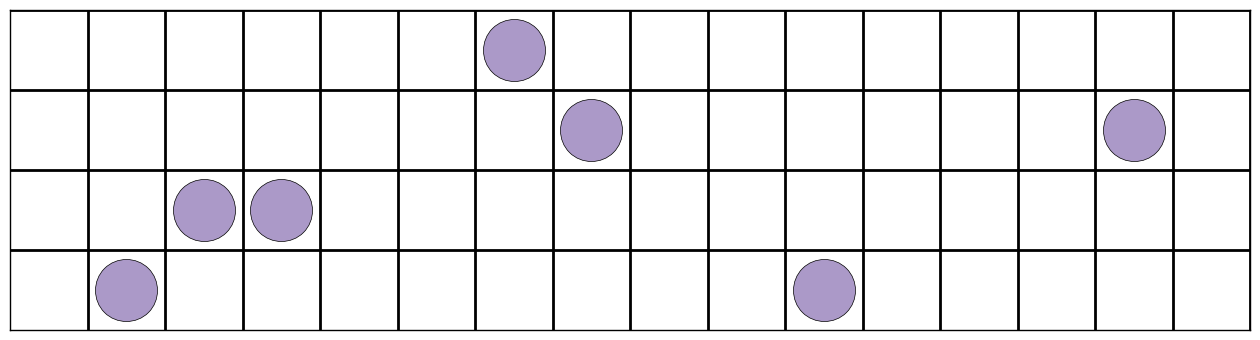}};

  \node[inner sep=0pt] (russell) at (3,-0.5) {$=$};

  \node[inner sep=0pt,label=below:{$\mathbf{a}_0$}] (russell) at (3.9,-0.5) {\includegraphics[height=40px]{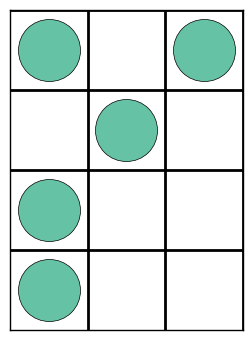}};
  \node[inner sep=0pt] (russell) at (5,-0.5) {$\circledast$};
  \node[inner sep=0pt,label=below:{$\mathbf{s}_0$}] (russell) at (7.7,-0.5) {\includegraphics[height=10px]{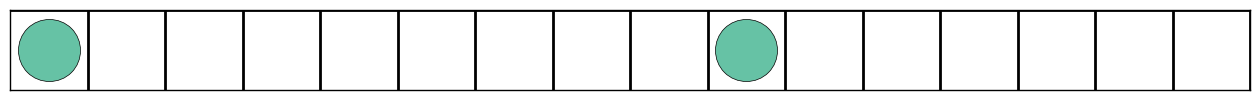}};

  \node[inner sep=0pt] (russell) at (6,-1.5) {$+$};

  \node[inner sep=0pt,label=below:{$\mathbf{a}_1$}] (russell) at (3.9,-2.5) {\includegraphics[height=40px]{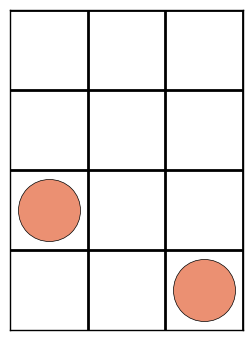}};
  \node[inner sep=0pt] (russell) at (5,-2.5) {$\circledast$};
  \node[inner sep=0pt,label=below:{$\mathbf{s}_1$}] (russell) at (7.7,-2.5) {\includegraphics[height=10px]{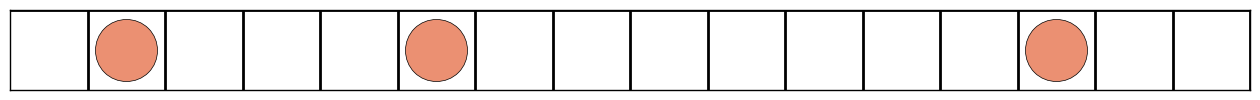}};

  \node[inner sep=0pt] (russell) at (6,-3.5) {$+$};

  \node[inner sep=0pt,label=below:{$\mathbf{N}$}] (russell) at (6,-4.5) {\includegraphics[height=40px]{assembly-noise.png}};

  \end{tikzpicture}
  \caption[Convolutional coding]{\textbf{Convolutional coding:} In this example the raw signal $\mathbf{Y}$ is an additive mixture of two motifs $\mathbf{a}_0$ and $\mathbf{a}_1$ (which have been highlighted with different colors) convolved with their activities $\mathbf{s}_0$ and $\mathbf{s}_1$ resulting in the reconstruction $\mathbf{X}$. Additional background noise $\mathbf{N}$ consists of non-repeating spikes. Both the motifs and their activities are learned simultaneously.}
  \label{fig:schematic}
\end{figure}

Let $\mathbf{Y} \in\mathbb{R}_+^{n\times m} $ be a matrix whose $n$ rows represent individual neurons with their spiking activity binned to $m$ columns. We assume that this raw signal is an additive mixture of $l$ of motifs $\mathbf{a}_i \in\mathbb{R}_+^{n\times\tau}$ with temporal length $\tau$ convolved with a sparse activity signal $\mathbf{s}_i$ plus noise (see figure \ref{fig:schematic}). 

We address the unsupervised problem of simultaneously estimating both the coefficients making up the motifs $\mathbf{a}_i$ and their activities $\mathbf{s}_i$. To this end, we propose to solve the following optimization problem:
\begin{align}
   &\min_{\mathbf{a}, \mathbf{s}} \left\| \mathbf{Y} - \sum_i^l \mathbf{s}_i \circledast \mathbf{a}_i \right\|_F^2 + \alpha \sum_i^{l} \|\mathbf{s}_i\|_0 + \beta \sum_i^l \| \mathbf{a}_i \|_1 \label{eq:ensembles}
\end{align}

The \lnorm{0} norm is chosen for $\mathbf{s}$ since \textcite{andilla2014sparse} have successfully used it to learn spike trains of neurons. For the ensembles themselves the \lnorm{1} norm is used to enforce only few non-zero coefficients \cite{Zou05regularizationand}.

This problem is non-convex in general but can be approximated by initializing $\mathbf{s}$ to random noise and using a block coordinate descent strategy \cite[Section 2.7]{bertsekas1999nonlinear} to alternatingly optimize for the two variables. The activities $\mathbf{s}$ are inferred using convolution matching pursuit \cite{Szlam_10_convolutional,mallat_93_matching,Elad} and the ensembles themselves using LASSO regression with non-negativity constraints \cite{tibshirani1996regression} by transforming the convolution with $\mathbf{s}_i$ to a linear set of equations using Toeplitz matrices $\tilde{\mathbf{s}_i}$ with $\tilde{\mathbf{s}}_{i,j,k} = \tilde{\mathbf{s}}_{i,j+1,k+1} = \mathbf{s}_{i,j-k}$ for $j \geq k$ and $\tilde{\mathbf{s}}_{i,j,k} = 0$ for $j < k$ (where $i$ denotes the $i$th matrix with element indices $j$ and $k$) which are then stacked next to each other: \cite{hansen2002deconvolution,Zou05regularizationand}:
        \begin{align}
          & \min_{\mathbf{a}} \left\| \vectorize(\mathbf{Y}) - \sum_i^l \mathbf{\tilde s}_i \cdot \vectorize(\mathbf{a}_i) \right\|_2^2 + \beta \sum_i^l \| \mathbf{a}_i \|_1 \nonumber \\
          = & \min_{\mathbf{a}} \left\| \underbrace{\vectorize(\mathbf{Y})}_{b \in \mathbb{R}^{mn}} -
            \underbrace{\begin{bmatrix}
              \mathbf{\tilde s}_0 & ... & \mathbf{\tilde s}_l
            \end{bmatrix}}_{A \in \mathbb{R}^{mn \times ln\tau}}
            \underbrace{\begin{bmatrix}
              \vectorize(\mathbf{a}_0) \\ ... \\ \vectorize(\mathbf{a}_l)
            \end{bmatrix}}_{x \in \mathbb{R}^{ln\tau}}
          \right\|_2^2 + \beta \sum_i^l \| \mathbf{a}_i \|_1
        \label{eq:enet}
        \end{align}

\newcommand{\myshiftheight}{50px}
\begin{figure}[h]
    \centering
    \subcaptionbox{Ground truth ensemble\label{fig:shift-gt}}[.29\linewidth]{\includegraphics[height=\myshiftheight]{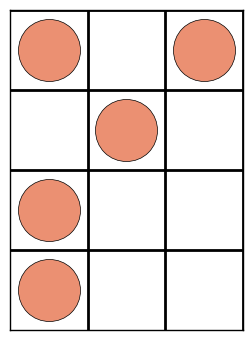}}
    \subcaptionbox{Possible wrong ensembles after a single iteration\label{fig:shift-wrong}}[.29\linewidth]{\includegraphics[height=\myshiftheight]{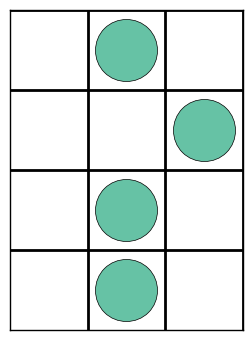} \includegraphics[height=\myshiftheight]{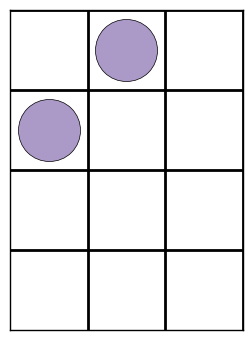}}
    \subcaptionbox{Equivalent padded ensembles\label{fig:shift-better}}[.4\linewidth]{\includegraphics[height=\myshiftheight]{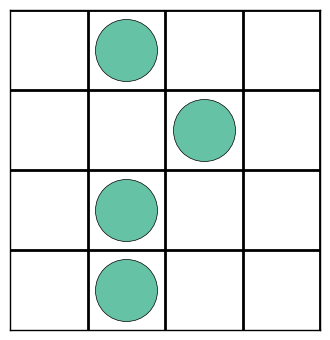} \includegraphics[height=\myshiftheight]{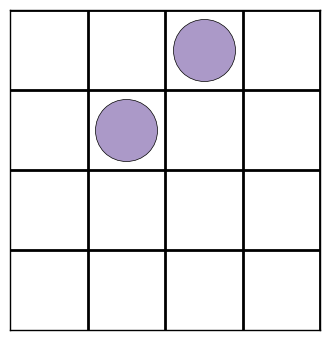}}
    \caption[Ground truth ensemble vs. learned, shifted ensemble]{\textbf{Ground truth ensemble and learned state after a single iteration}: This figure shows four neurons forming a motif over three bins. After a single iteration parts of the motif can be missing which is solved by increasing the ensemble length and centering the motifs after each iteration.}
\end{figure}

Special care has to be taken to avoid missing parts of the motif due to the originally identified positions. Consider the ground truth ensemble seen in figure \ref{fig:shift-gt}. After a single iteration the learned ensemble could be any of the two wrong possibilities seen in figure \ref{fig:shift-wrong}. While the learned motif does indeed occur in real data it is not complete and can never be completed since there is no more space on the left or right to identify the missing associations. To overcome this problem the vectors $\mathbf{a}$ have to be chosen larger than required and centered after each iteration when possible: When there are enough empty rows on either side the whole motif can be shifted before the new ensemble activities $\mathbf{s}$ are identified. This does not increase the reconstruction error since the activities will also just be shifted by the same amount. When new coefficients $\mathbf{a}$ are learned in the next iteration there now is enough space to also capture the previously missed associations
 (see figure \ref{fig:shift-better} which can be completed in the next iteration).

\section{Results on synthetic data}

Since ground truth datasets are in general not available different synthetic data sets consisting of fifty neurons observed over one thousand time bins have been simulated to compare our method to existing work. A subset of the neurons is randomly assigned to belong to a single ensembles, others  to multiple ensembles and the rest are not part of any ensemble and fire completely on their own. The ensemble activity itself is modeled as a Poisson process with a randomly chosen mean \cite{lopes2013detecting} and a refractory period of at least the length of the ensemble itself. Additionally spurious spikes of single neurons are added to simulate neurons firing out of sync. The percentile of neurons belonging to multiple ensembles, the fraction of spurious spikes and the temporal lengths of the ensembles have been varied to create different test cases.

For PCA and ICA based methods the number of ensembles is estimated using the Marchenko-Pastur eigenvalue distribution \cite{lopes2013detecting}. The sparsity parameter in O'Grady's sparse convolutive non-negative matrix factorization that resulted in the best performance was experimentally chosen \cite{o2006convolutive}. Additionally the normalized correlation between the spike trains of each possible neuron-neuron combination has been calculated. If the correlation between neuron $i$ and neuron $j$ is higher than the correlation between $i$ and any other neuron those two are assumed to be connected within an ensemble.

\begin{figure}[t!]
	\subcaptionbox{Spike matrix\label{fig:synth-example-spikes}}{\includegraphics[height=150px]{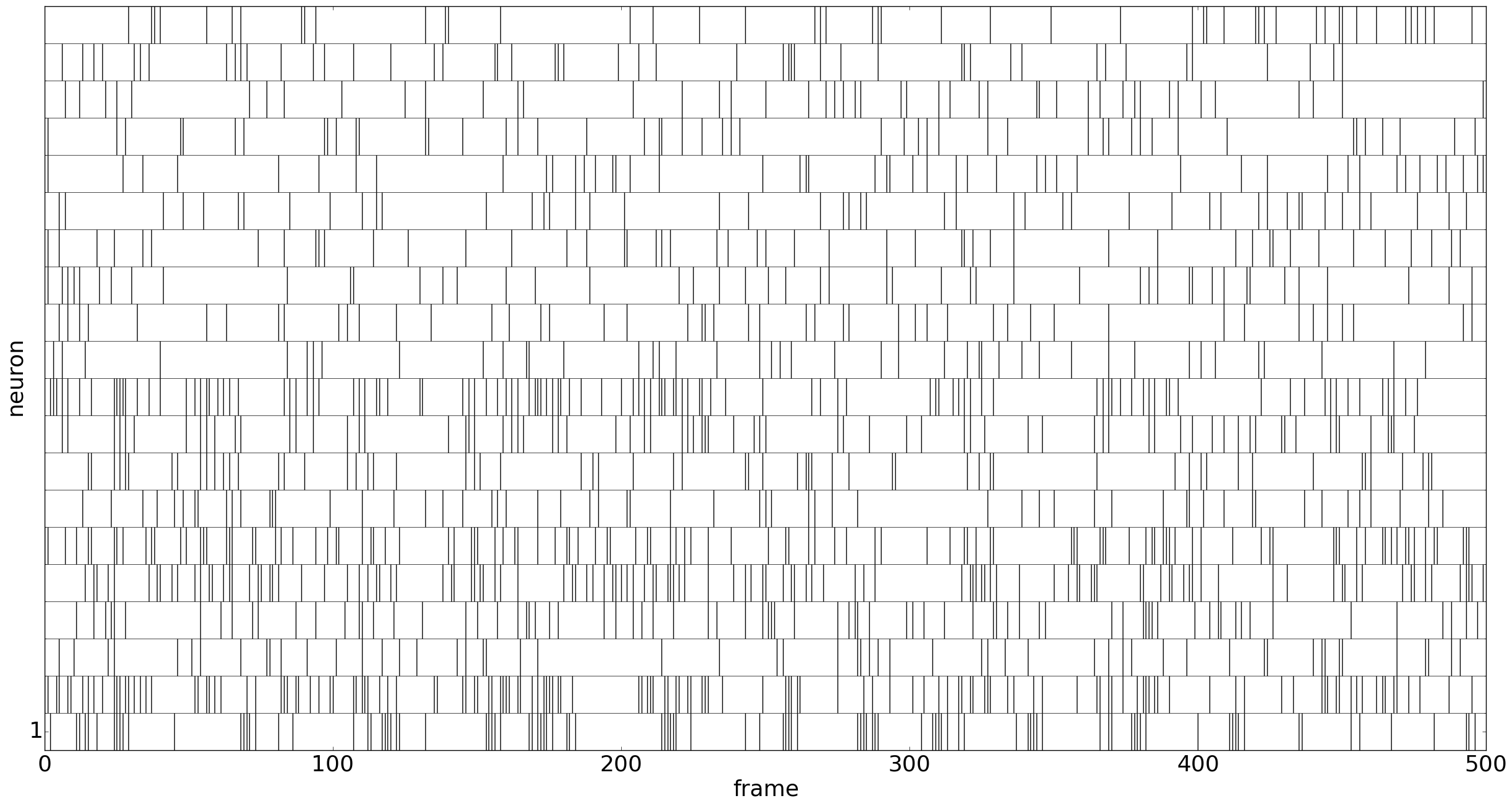}}
	\subcaptionbox{Ground truth motifs\label{fig:synth-example-ensemble-gt}}{\includegraphics[height=90px]{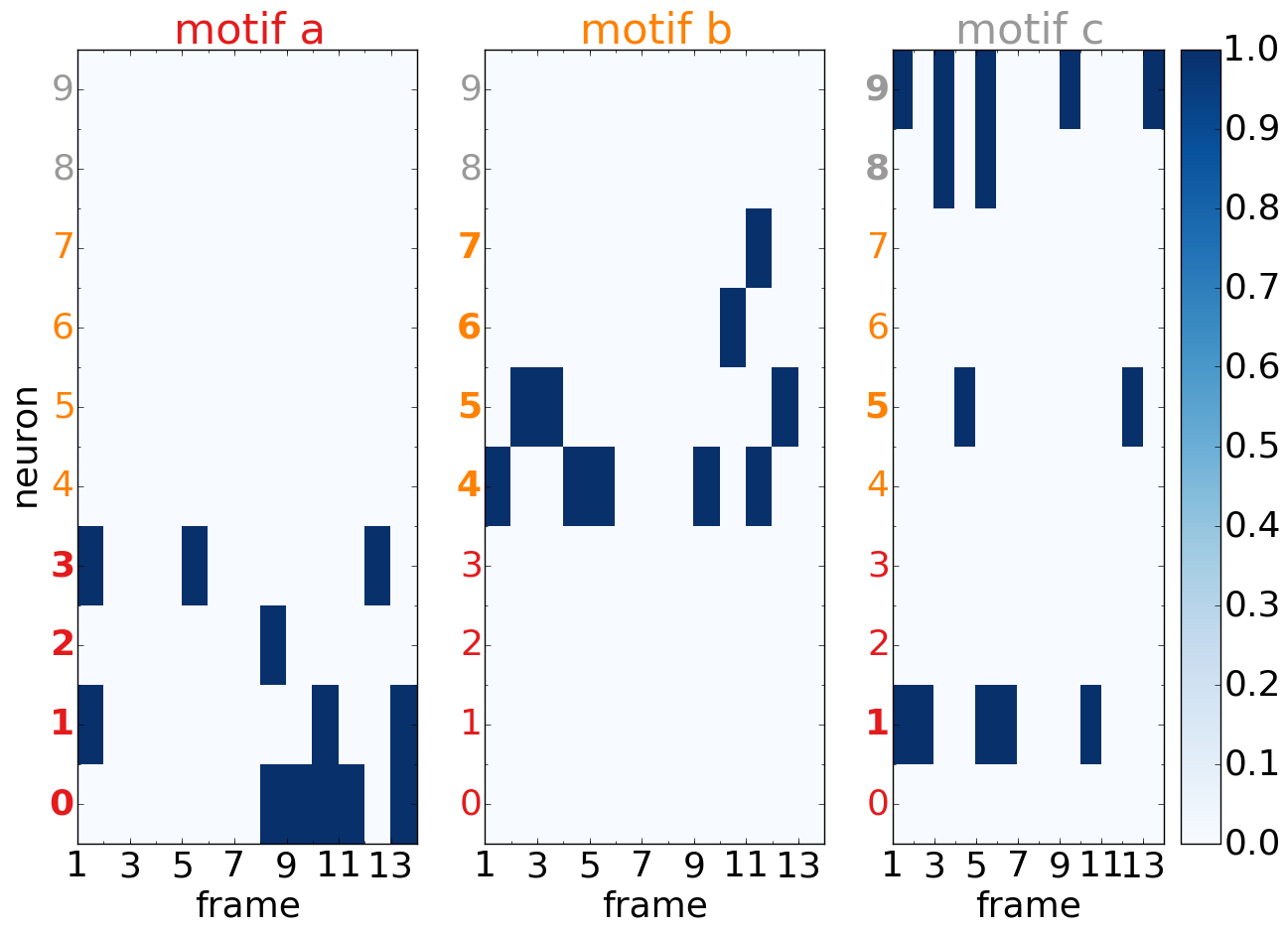}}
	\subcaptionbox{Learned motifs (first trial)\label{fig:synth-example-ensemble-learned}}{\includegraphics[height=90px]{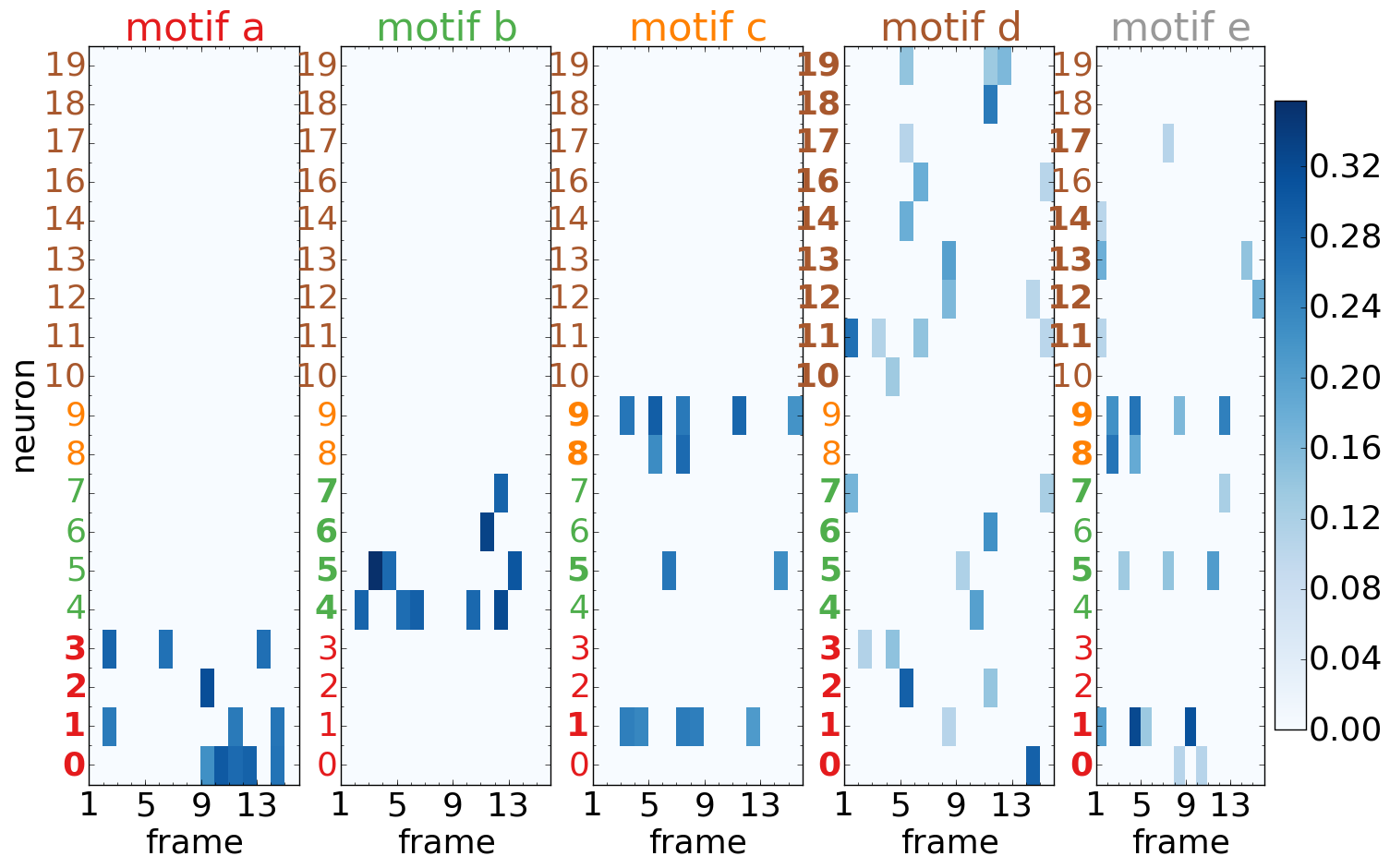}}
	\subcaptionbox{Learned motifs (second trial)\label{fig:synth-example-ensemble-learned2}}{\includegraphics[height=90px]{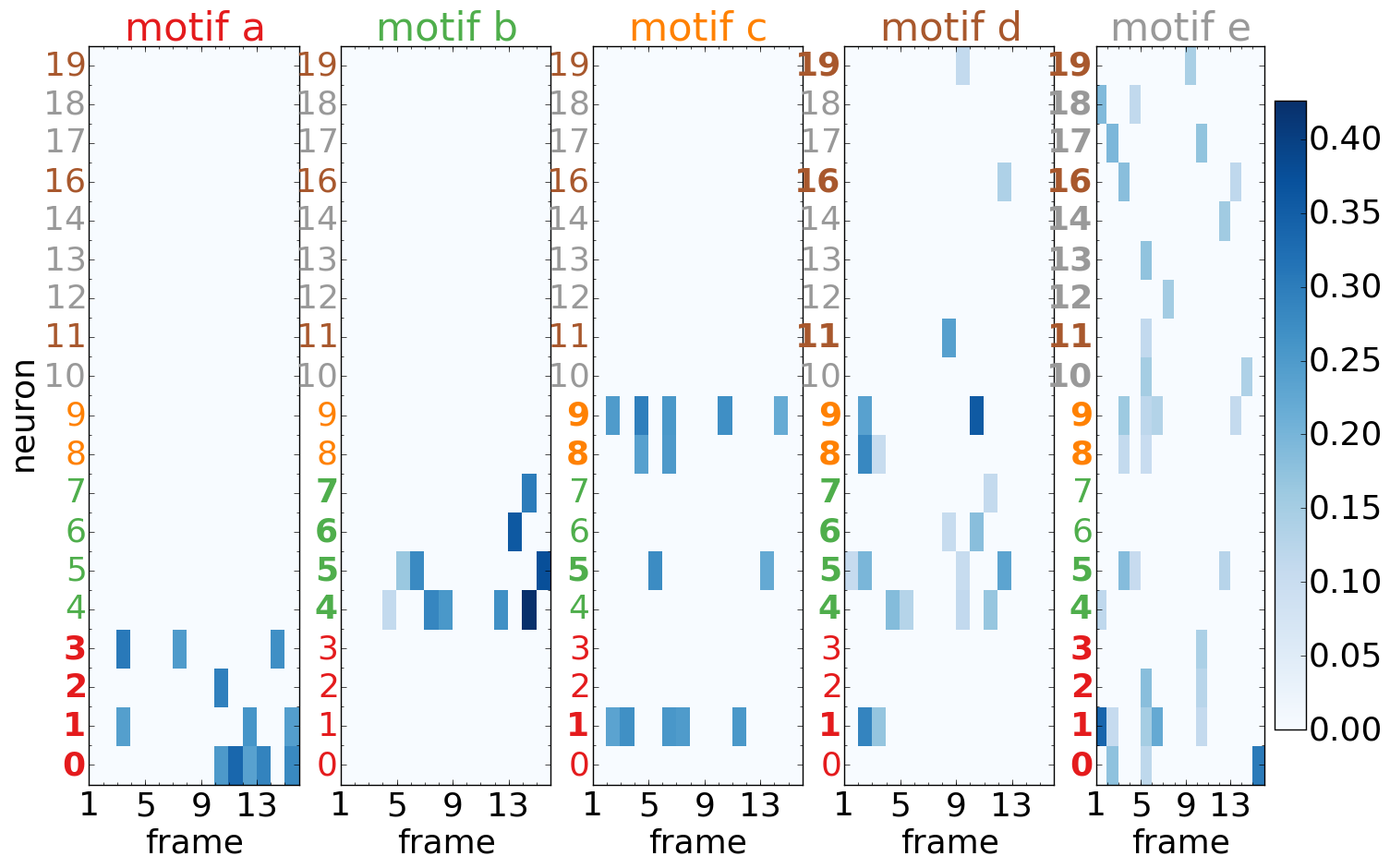}}
	\subcaptionbox{Learned component (PCA)\label{fig:synth-example-ensemble-learned-pca}}{\includegraphics[height=90px]{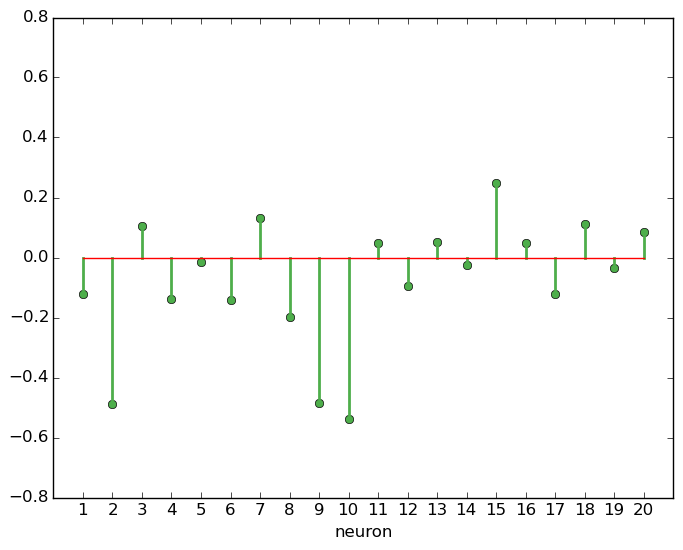}}
	\subcaptionbox{Learned component (ICA)\label{fig:synth-example-ensemble-learned-ica}}{\includegraphics[height=90px]{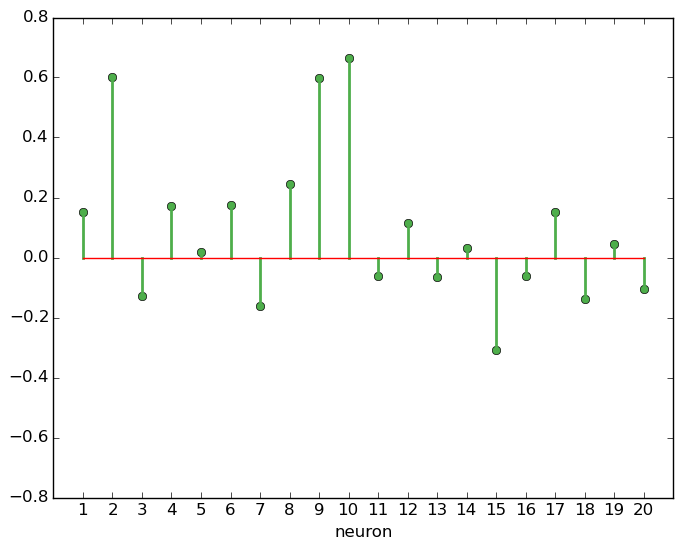}}
	\subcaptionbox{Learned motifs (scNMF)\label{fig:synth-example-ensemble-learned-scnmf}}{\includegraphics[height=90px]{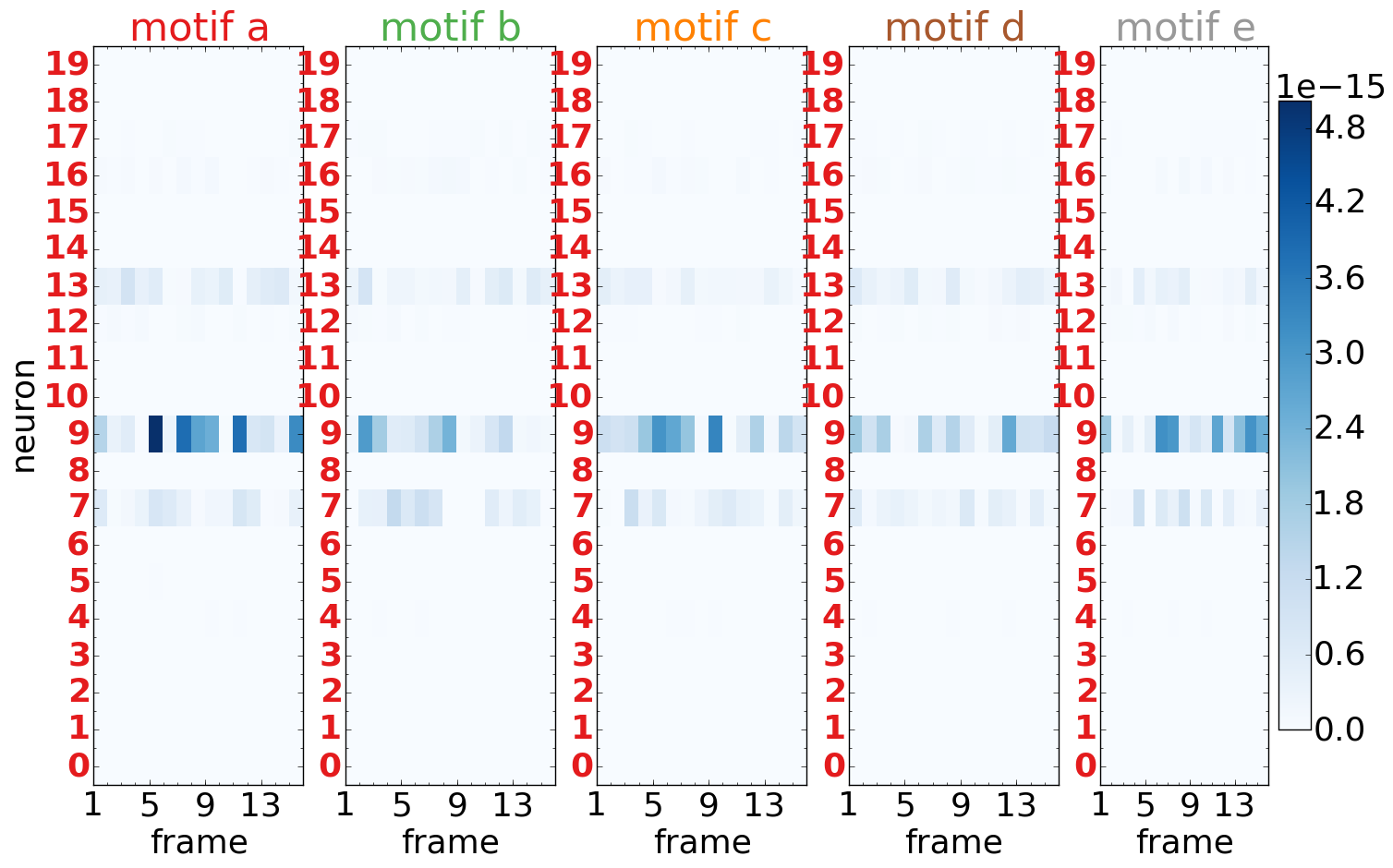}}

    \bfcaption{Results on a synthetic dataset}{\subref{fig:synth-example-spikes} shows a synthetic spike matrix with twenty neurons and 500 frames. Each neuron spikes in fifty randomly distributed bins. \subref{fig:synth-example-ensemble-gt} shows the three motifs made up by the first ten neurons. The time between two activations of a motif has been modeled as Poisson distributed with a mean distance of twenty frames. By running our algorithm with two different random initial states the motifs seen in \subref{fig:synth-example-ensemble-learned} and \subref{fig:synth-example-ensemble-learned2} are learned. \subref{fig:synth-example-ensemble-learned-pca}, \subref{fig:synth-example-ensemble-learned-ica} and \subref{fig:synth-example-ensemble-learned-scnmf} show that no other method is able to learn a single true motif.}
    \label{fig:synth-example}

\end{figure}

A simplified example dataset with twenty neurons, three temporal motifs and fifty spurious spikes per neuron can be seen in figure \ref{fig:synth-example}. When running our method with two different random initial states to identify five motifs with a temporal length of fifteen frames for ten iterations all three original motifs are learned successfully (figure \ref{fig:synth-example-ensemble-learned} and \ref{fig:synth-example-ensemble-learned2}; the motifs have been sorted manually to match up with the ground truth). The two spurious motifs change depending on the random initial state while the true motifs always appear in the results. Neither PCA, ICA or scNMF are able to learn any true motif.

To evaluate the different methods a neuron association matrix is calculated from the learned ensembles by choosing a threshold above which neurons are assumed to belong to an ensemble and compared to the ground truth association matrix. The functional association between neurons has been used as an indicator of performance in previous work \cite{lopes2011neuronal,carrillo2015endogenous,Billeh201492} but does not take the identification of the correct temporal motifs into account. We still chose this method since it works without limitations for synchronous motifs and also allows comparisons for the more complex cases.

\newcommand{\myrocheight}{150px}
\begin{figure}[t!]
    \caption{\textbf{ROC curves for different temporal motif lengths}: For each of the three experiments twenty different datasets (with different noise levels and amount of neurons shared between ensembles) were generated and all methods were run ten times with different random initialization. We show the mean ROC curve and its standard deviation averaged over all trial done on different synthetic datasets.}

	\subcaptionbox{$\tau = 1$\label{fig:synth-roc-1}}{\includegraphics[height=\myrocheight]{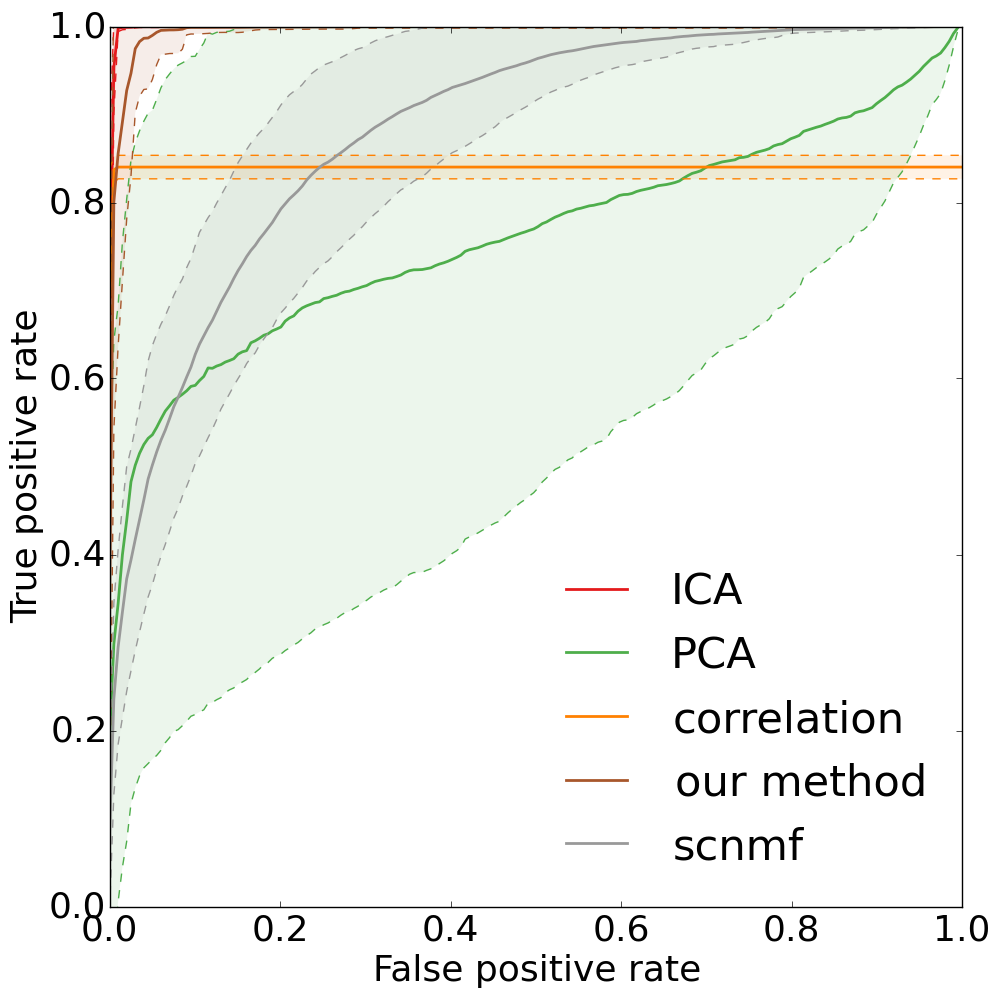}}
	\subcaptionbox{$\tau = 7$\label{fig:synth-roc-7}}{\includegraphics[height=\myrocheight]{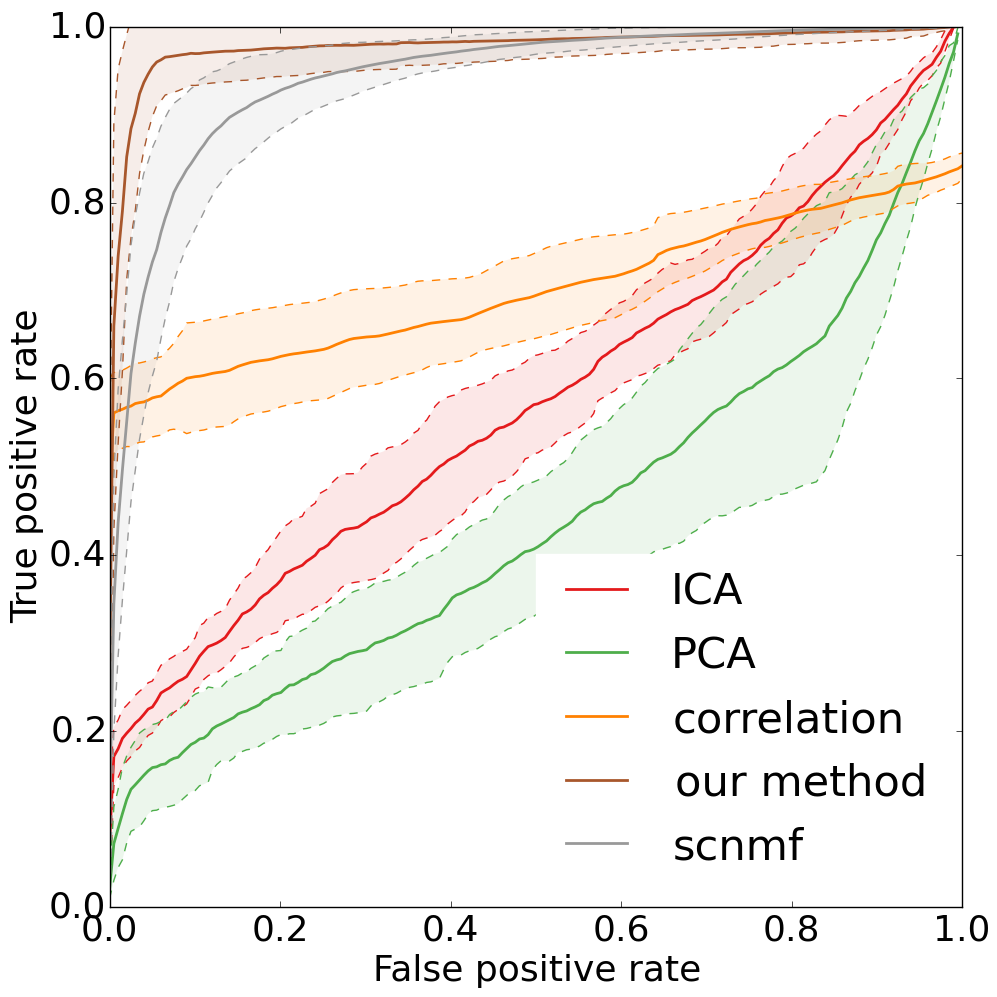}}
	\subcaptionbox{$\tau = 21$\label{fig:synth-roc-21}}{\includegraphics[height=\myrocheight]{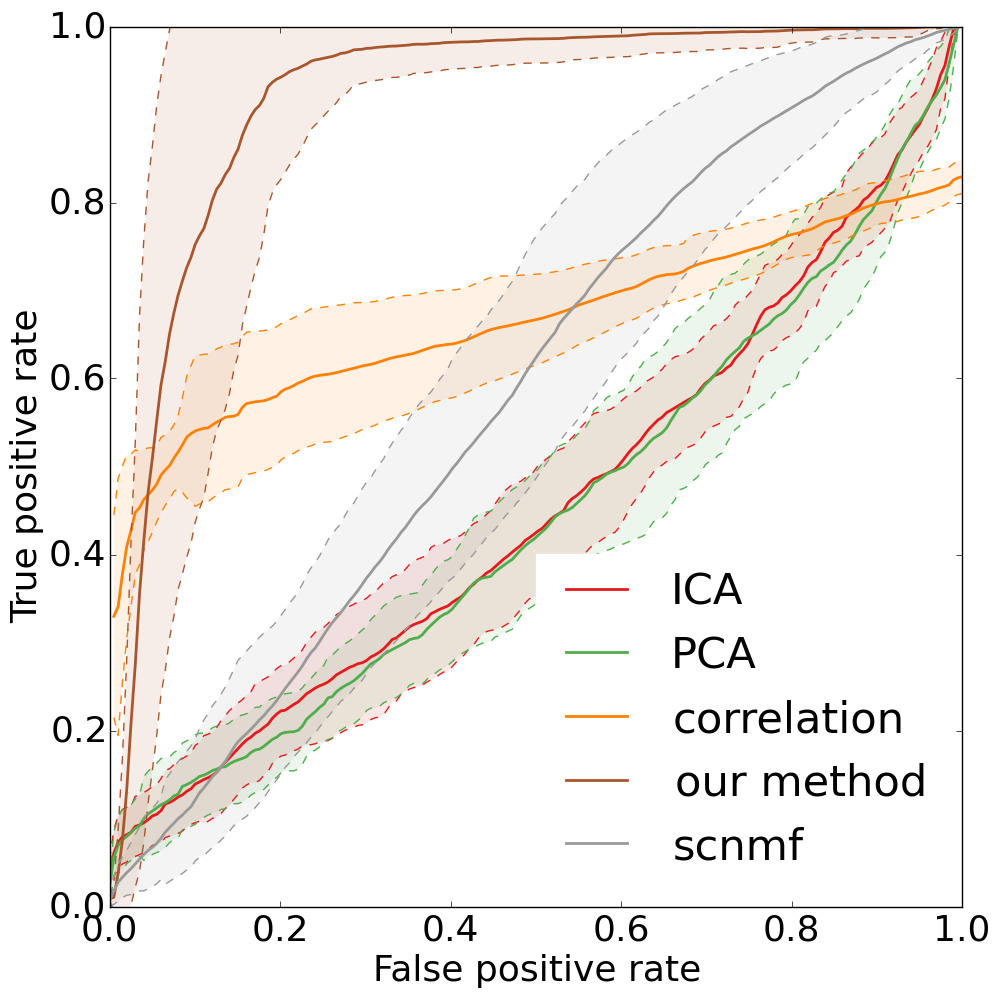}}

    \label{fig:synth-roc}
\end{figure}

We plotted three different ROC curves for the different temporal lengths $\tau = 1,7 \text{ and } 21$ in figure \ref{fig:synth-roc}. In the synchronous case (i.e. $\tau = 1$, figure \ref{fig:synth-roc-1}) our proposed method performs as good as the best competitor. As expected PCA performance shows a huge variance since some of the datasets contain neurons shared between multiple ensembles and since extracting actual neuron-ensemble assignments is not always possible \cite{lopes2011neuronal,lopes2013detecting}.

When temporal structure is introduced we are still able to identify associations between neurons with very high accuracy. Sparse convolutional matrix factorization is able to identify associations for short temporal motifs ($\tau = 7$, figure \ref{fig:synth-roc-7}) but only we are able to accurately recover most associations in long motifs ($\tau = 21$, figure \ref{fig:synth-roc-21}).

\begin{figure}[t!]
	\subcaptionbox{Ground truth association matrix\label{fig:synth-stability-gt}}[.29\linewidth]{\includegraphics[height=100px]{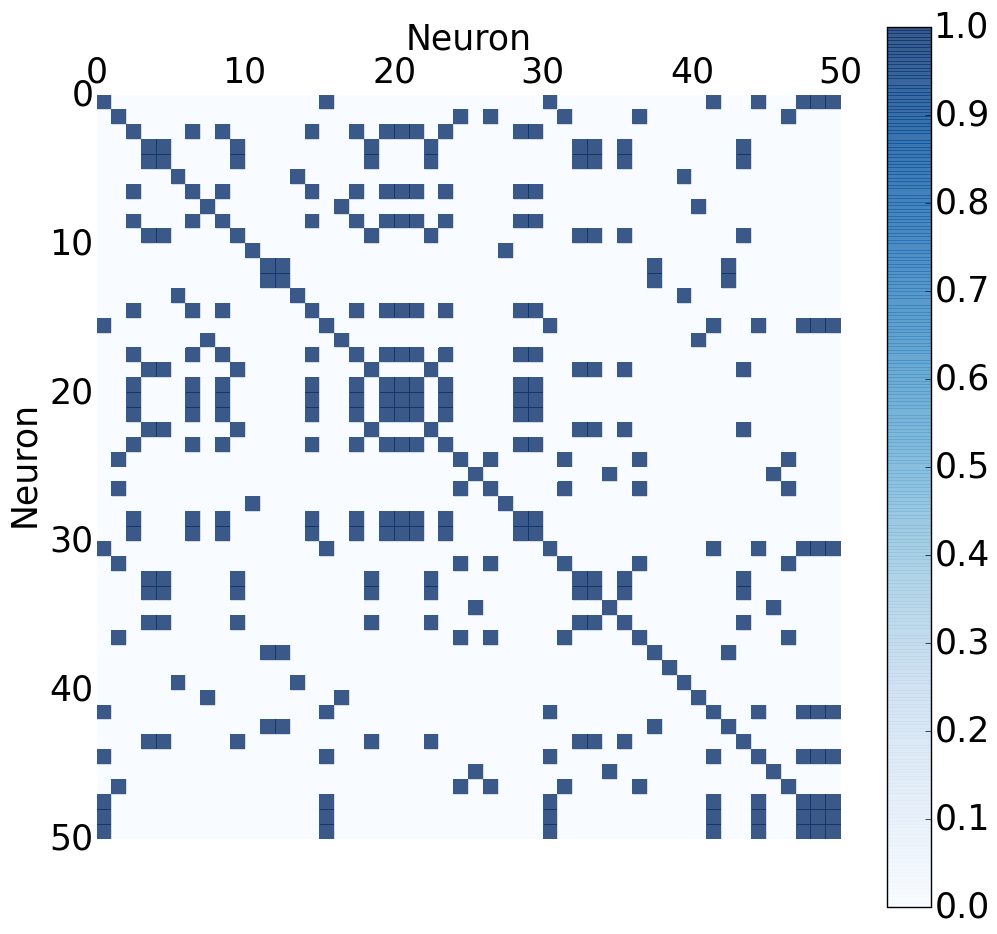}}
	\subcaptionbox{Learned association matrix\label{fig:synth-stability-learned}}[.29\linewidth]{\includegraphics[height=100px]{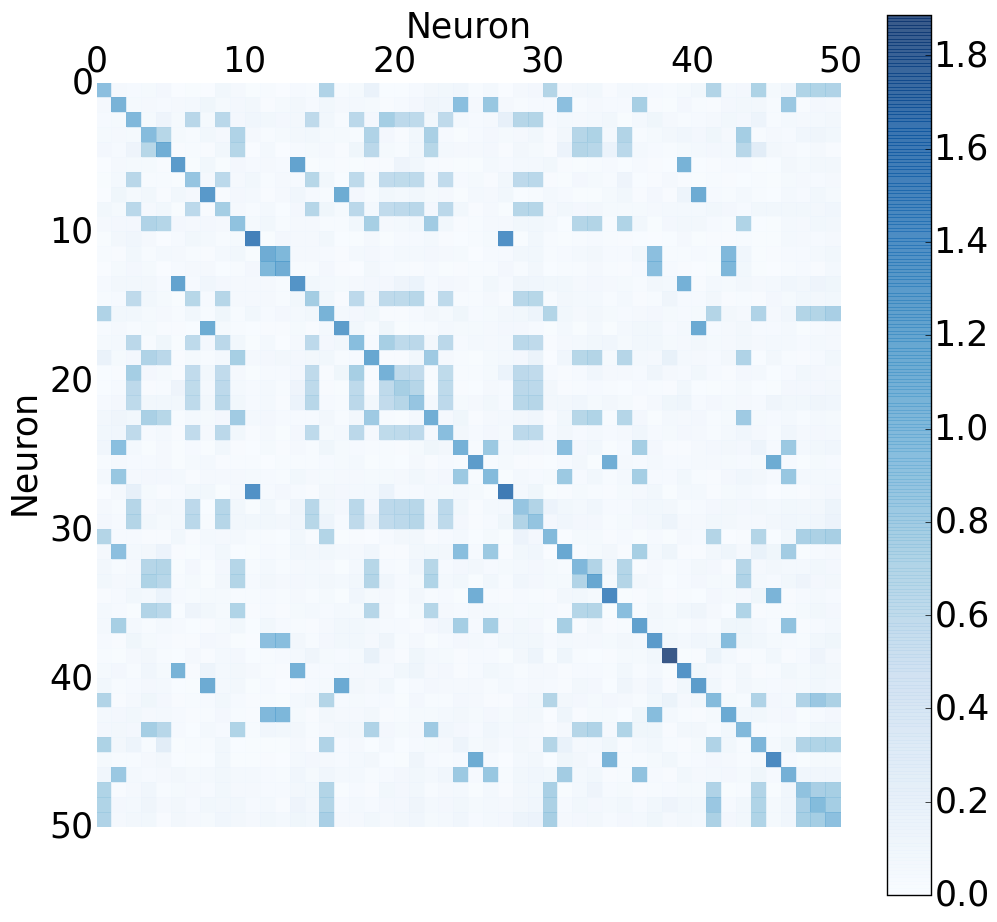}}
	\subcaptionbox{Learned association matrix on shuffled data\label{fig:synth-stability-shuffle}}[.29\linewidth]{\includegraphics[height=100px]{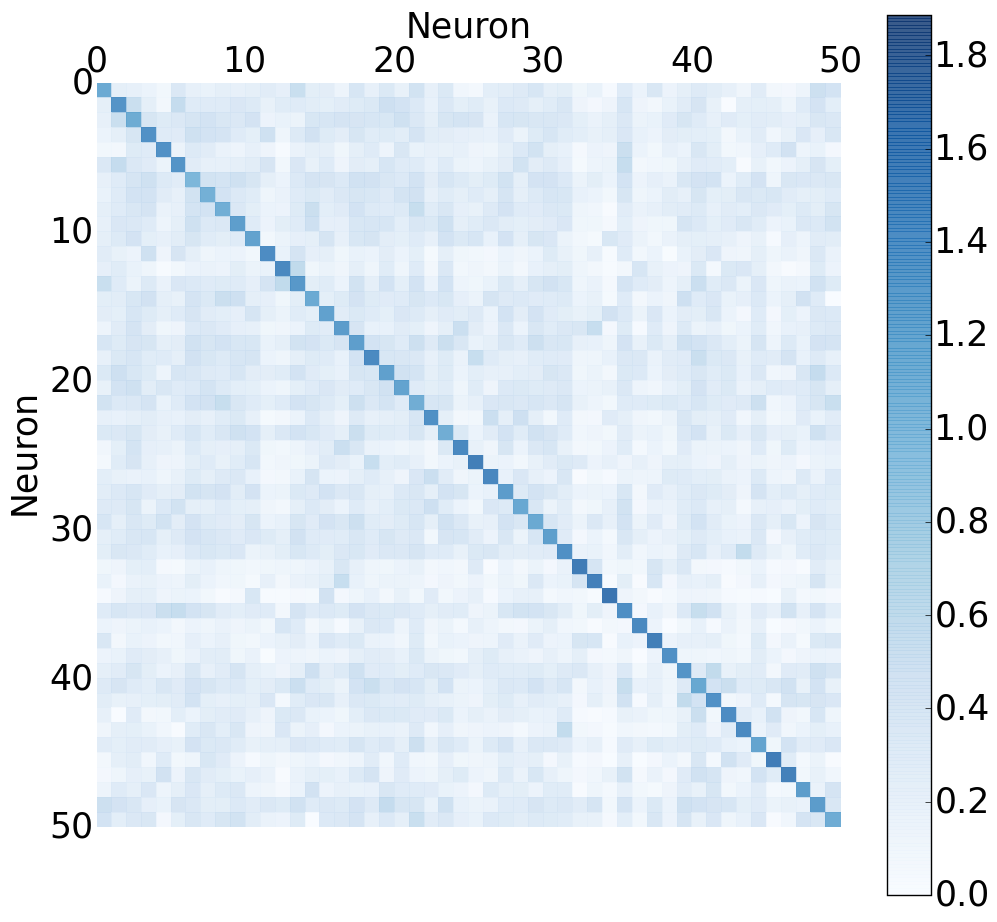}}

	\subcaptionbox{Association matrix values\label{fig:synth-stability-thres}}[.59\linewidth]{\includegraphics[height=150px]{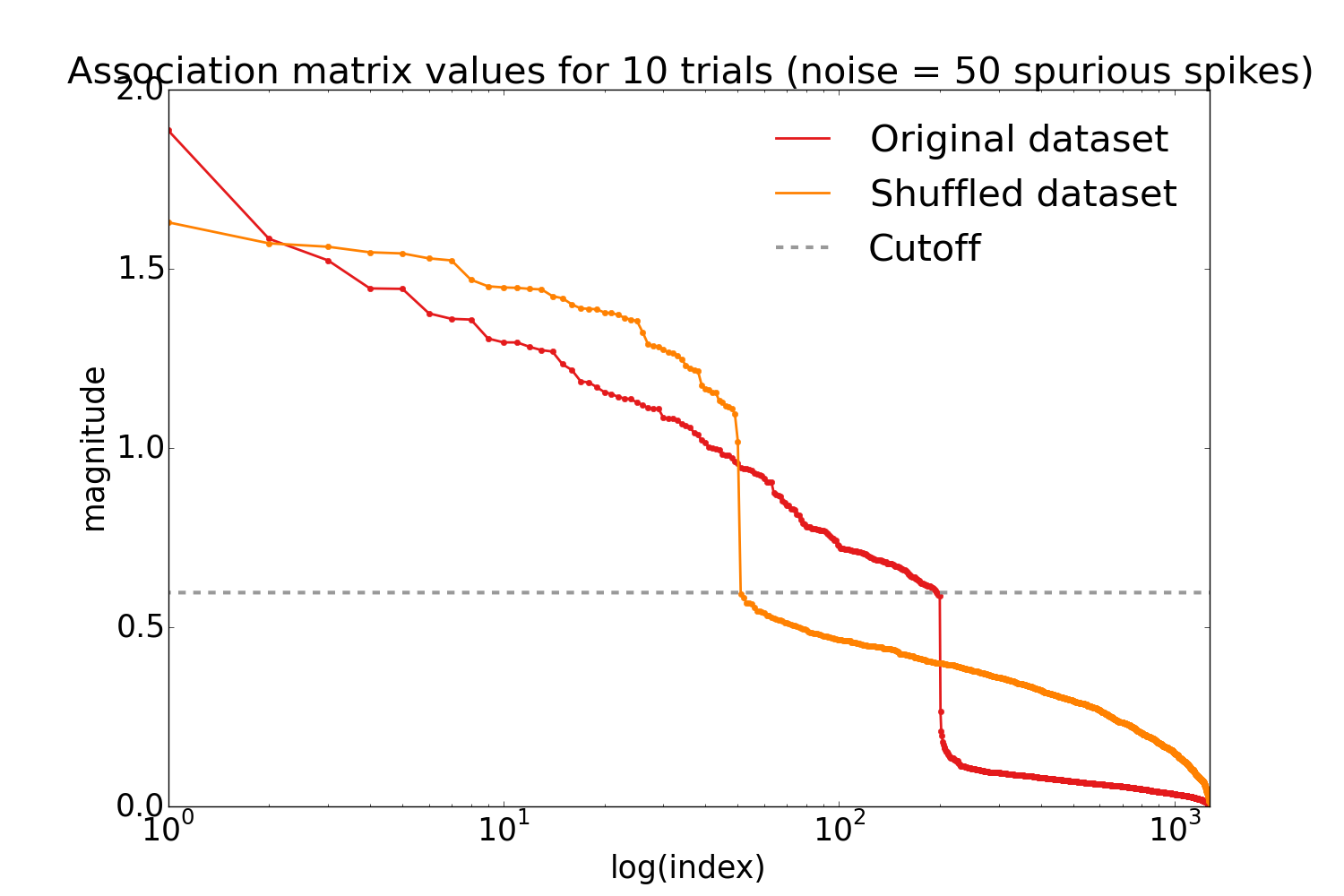}}
	\subcaptionbox{Missing associations after thresholding in the unshuffled data\label{fig:synth-stability-shuffle-thres}}[.29\linewidth]{\includegraphics[height=100px]{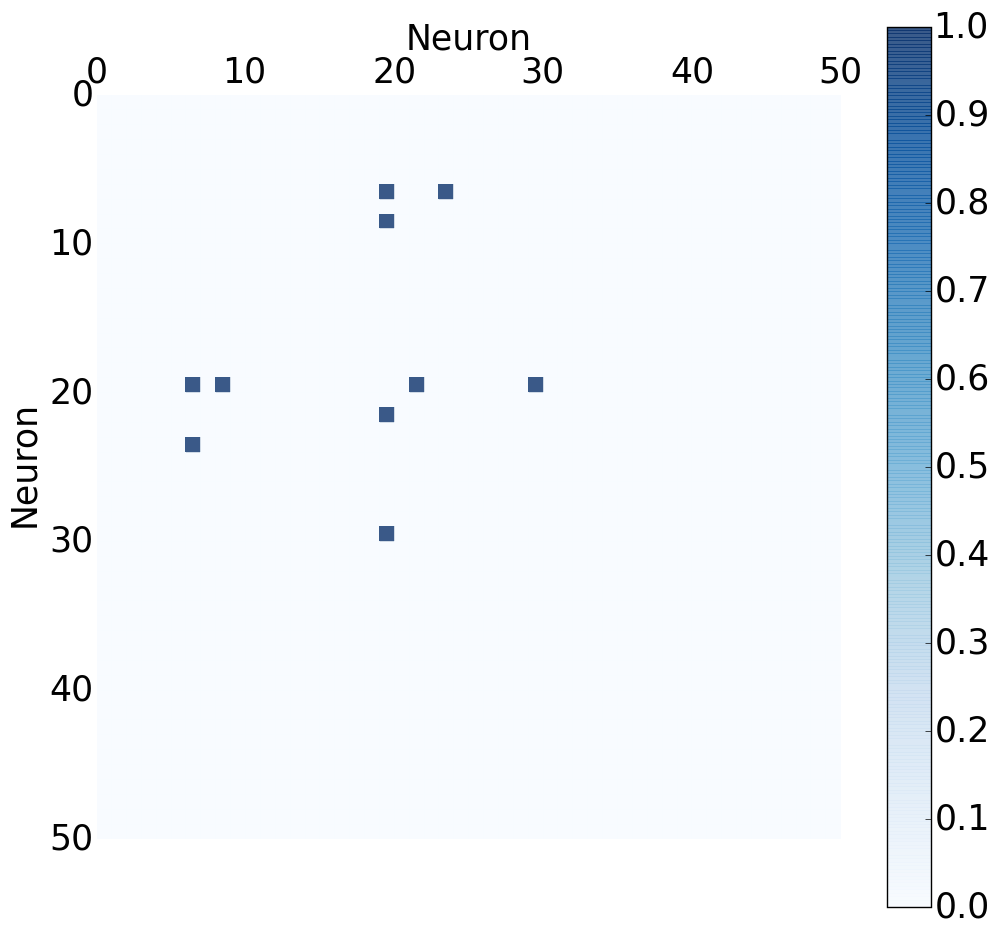}}


    \bfcaption{Stability analysis of the proposed method}{A synthetic dataset consisting of ten synchronous motifs with fifty neurons in total was generated. The spike vector for each neuron is shuffled independently to destroy all correlations. Our proposed methods is then run with ten different initial states on both datasets and the resulting average association matrices are computed. A threshold corresponding to a significant change in slope of the shuffled association matrix is chosen and applied to the unshuffled matrix. Using this method only five neuron-neuron associations are not identified.}
    \label{fig:synth-stability}

\end{figure}

In order to study the stability of our results and the effects of the random initialization of the spike trains we use a non parameterized tests similar to the ones used by other methods to estimate the number of ensembles \cite{lopes2013detecting,carrillo2015endogenous}. Starting with the same generated dataset consisting of ensembles with no temporal motifs, i.e. subsets of neurons firing synchronously, we shuffle the spike trains of each neuron independently in order to preserve the spike distribution while destroying correlations between individual neurons. We then run our method ten times on the original and the shuffled dataset with different random initializations and compute the average neuron association matrix. Then all unique values inside both matrices are sorted and plotted in figure \ref{fig:synth-stability-thres}. The point after a significant change in slope in the shuffled matrix is chosen as a threshold on the original matrix. Only five connections are discarded using this procedure.

\section{Conclusion}
Especially in the past two decades where the number of neurons that can be recorded simultaneously was drastically increased \cite{buzsaki2004large,stevenson2011advances,ahrens2013whole} many studies have been done to identify neurons firing in patterns to form motifs as originally suggested by Hebb \cite{hebb1949organization}. While many algorithms have been developed almost all of those are limited to the identification of synchronous motifs only \cite{chapin1999principal,lopes2013detecting,diego_13_learning,Billeh201492,carrillo2015endogenous}.

We have presented a new mathematical method for the identification of motifs that is not limited to synchronous activity. Our method leverages sparsity constraints on the activity and the motifs themselves to allow a simple and elegant formulation that is able to also learn motifs with temporal structure. The proposed algorithm extends convolutional coding approaches, which have previously already been successful in recovering spike trains from calcium fluorescence recordings \cite{andilla2014sparse,pnevmatikakis_13_rank,pnevmatikakis_13_sparse,Vogelstein2009b} and the identification of repeating patterns in audio data \cite{o2006convolutive}, with a novel optimization approach to allow modeling of interactions between neurons.

Results on simulated datasets show that the proposed method outperforms others especially when identifying long motifs (figure \ref{fig:synth-roc}).

We hope that these contributions allow to study more complex neuronal firing patterns and help to further the understanding of functional ensembles within the brain as originally suggested by Hebb.

\subsection*{Acknowledgments}
This work is a project of the SFB 1134 “Functional Ensembles” funded by the Deutsche Forschungsgemeinschaft (DFG).

\printbibliography

\end{document}